# Some New Views on Product Space and Related Diversification

Önder Nomaler and Bart Verspagen

UNU-MERIT

**Abstract**:

We aim to contribute to the literature on product space and diversification by proposing a number of extensions of the current literature: (1) we propose that the alternative but related idea of a country space also has empirical and theoretical appeal; (2) we argue that the loss of comparative advantage should be an integral part of (testing the empirical relevance of) the product space idea; (3) we propose several new indicators for measuring relatedness in product space; and (4) we propose a non-parametric statistical test based on bootstrapping to test the empirical relevance of the product space idea.

## 1. Introduction

Launched by a series of papers by Hidalgo, Hausmann and Klinger (Hausmann and Klinger 2006; 2007; Hidalgo et al. 2007), the idea of a product space in which country's specializations are path dependent through the process of related diversification has gained popularity for explaining the trade performance of countries and regions. The basic underlying idea, although never explicitly modeled in terms of micro foundations, is that productive capabilities of countries (or regions) are developed "locally", i.e., by incremental changes, and that this creates a "relatedness" between the products that are produced by these specific capabilities (e.g., Hidalgo et al., 2007). This relatedness can be observed as a "product space" in which individual countries tend "produce" (which is taken to mean "export with comparative advantage", see below) combinations of related products more often than combinations of unrelated products. New specializations (gains of new comparative advantages, or "innovations") will then tend to occur in products that are related to the existing specialization pattern of the country or region.

Our aim in this paper is to extend and elaborate on the idea of product space in several ways. In particular, we aim to develop the idea in four different ways: (1) we propose that the alternative but related idea of a country space also has empirical and theoretical appeal; (2) we argue that the loss of comparative advantage should be an integral part of (testing the empirical relevance of) the product space idea; (3) we propose several new indicators for measuring relatedness in product space, including one measure that decomposes the probability of gaining new specializations into related and unrelated diversification, and which, thereby, provides a more precise measure of path-dependency in product space; and (4) we propose a non-parametric statistical test based on bootstrapping to test the empirical relevance of the product space idea both at the country- and product-level, and including both gains and losses of comparative advantage. Before introducing these new ideas, we will first provide a brief overview of the existing literature, and use this to position the proposed extensions that are developed in the next sections.

The product space idea has profound implications for development. In a nutshell, it means that developed and developing nations tend to produce different kinds of products: development is a process of (among other things) creating, adapting and adopting new productive capabilities that are associated with different parts of the product space. As argued by, e.g., Coniglio et al. (2021), this ties the product space literature to the literature that looks at the relationship between development and trade diversification, e.g., Imbs and Warcziarg (2003) and Hausmann and Rodrik (2003). Development then becomes associated with a move through product space, i.e., by the process of 'related' diversification.

However, as argued in more detail in Nomaler and Verspagen (2021), specialization and diversification are related but ultimately also different concepts, and to some extent they even contradict each other. If we define diversification as the increase in the number of products of sectors that a country or region exports with comparative advantage (this is the implicit definition that is most often used in the product space literature), then diversification clearly has limits, because a country (or region) cannot, by definition, be specialized in all products at the same time. Hence, if new specializations are added (diversification), we must also expect that at some point old specializations will vanish. This is clearly visible in the data that we will use below. In our dataset, 59 out of 155 countries have a higher number of losses than gains of comparative advantages over the six-year period 2012 – 2018. China, which is a country that is often seen as an example of development through diversification, did have a net gain of 143 comparative advantages over the period, but it lost 224 as well as gained 367. The loss of comparative advantage as an important part of the diversification process leads us to the second aim of the paper as formulated above, and also inspires our newly proposed indicators (third aim above).

The empirical evidence tends to show that the product space is an attractive idea. Hausman and Klinger (2007) introduce the so-called *density* measure for related diversification opportunities (we will discuss this measure in more detail below). Density is defined at the country (or region) – product level, and it measures how strongly the country or region is related to a specific product, through the existing specialization pattern of the country/region and the underlying product relatedness. Thus, if a country is already specialized in many products that are highly related to a specific product *j*, then the country will have a strong relatedness to product *j*, and if the country is not already specialized in *j*, then we may predict that diversification is likelier to lead to a new specialization in product *j*.

Hausman and Klinger (2007) use a *t*-test to show that the average of density of new specializations of countries is higher than the average density of unrealized diversification opportunities. They also employ OLS regressions with a comparative advantage indicator as the dependent variable and density as an explanatory variable to make the same point. The significantly positive sign of the density variable is taken as support for the product space idea of path dependence in specialization patterns.

The regression approach has also been followed in subsequent literature, e.g., Boschma and Capone (2015, 2016), Bahar et al. (2017), all performed similar regressions on a sample of countries and products. Alsonso and Martín (2019) apply the regression approach to Mexican regions. Guo and He (2015) apply the same regression approach to data on industrial employment data (including production for the domestic market). Coniglio et al. (2021), on the other hand, employ a non-parametric approach by comparing the distributions of density for cases where comparative advantage was actually gained and cases where comparative advantage could potentially be gained (i.e., comparative advantage did not exist). These studies overwhelmingly support the theoretical idea of product space and the path dependence of specializations that it predicts. But while most of the regression approaches use a "pooled" sample of countries (or regions) and products (or industries), Coniglio et al. (2021) also implement their test for individual countries, showing that some countries fit the product space idea, but others much less so.

The remainder of this paper is organized as follows. In the next section, we develop the theoretical side of our proposals by introducing a number of new indicators for relatedness of products and countries. In Section 3, we propose a non-parametric test for the predictive power of our indicators, introduce and describe the dataset that we use, and apply the empirical tests. This section also includes an empirical illustration of how one of our indicators can be decomposed to yield a version that is more targeted to path dependence of specializations. Section 4 summarizes our analysis and proposes some avenues for further development of the literature.

## 2. Measures

In this section, we develop a set of new measures for the relatedness between countries (or regions) and products. We start by summarizing the formalization of the density measure that is currently the most used measure. We then move on to propose new measures based on two of our proposals that we introduced above: the inclusion of the absence loss of comparative advantage, or, in other words, mutually exclusive specializations, in the analysis, and the idea of a "country space".

### *2.1. The density measure*

The most commonly used measure for a country's opportunities for diversification is referred to as *density*. This measure is based on the idea that a country's current specialization structure (partly) determines which products are likely targets for developing new comparative advantages (diversification). Diversification itself is measured through the notion of comparative advantage, denoted by the matrix $X$, with elements

$x_{ij} = 1$ if $\frac{E_{ij}/E_j}{E_i/E} \geq 1$ and $x_{ij} = 0$ otherwise,

where $E_{ij}$ denotes the value of exports of product $i$ by country $j$, and the absence of a subscript indicates summation over the relevant dimension. The matrix $X$ has dimensions $m$ x $n$, where $m$ is the number of products, and $n$ is the number of countries. Typically, $m \gg n$. We assume that each country exports at least one product, and each product is exported by at least one country (this ensures that all elements of $X$ are well-defined). A given $X$ matrix contains a total of $r = \sum_{\forall i \in \{1,2,..m\}} \sum_{\forall j \in \{1,2,..n\}} x_{ij}$ revealed comparative advantages.

The elements of $X$ are measures of so-called Revealed Comparative Advantage (RCA) of the column-country in the row-product. The density metric for related diversification (Klinger and Hausman, 2007) draws on the notion of conditional probability in RCA, computed on the basis of observed co-occurrences in $X$. The usual interpretation is that if conditional probability is high (low), the products are likely to share a high (low) degree of capabilities needed to export them with comparative advantage.

To formalize this, let $k_{pq}$ denote the number of countries that have comparative advantage both in product $q$ and in product $p$, and $s_p$ denote the number of countries which have comparative advantage in product $p$ ($s_p$ is what is usually called the 'ubiquity' of a product). Then $c_{qp} = k_{qp}/s_p$ denotes the probability that a country has comparative advantage in product $q$, conditional on the country having comparative advantage in product $p$. In matrix notation, these conditional probabilities are given by

$C = S^{-1}XX^T$,

where the superscript $T$ indicates a transposed matrix, and $S$ is the matrix with the corresponding row-sums of $X$ on the main diagonal and zeros elsewhere. Note that the diagonal of $S$ thus contains the ubiquity of respective products, i.e.,

$$s_{ij} = \begin{cases} \sum_{\forall k \in \{1,2,..n\}} x_{ik} & if\ i = j \\ 0 & otherwise \end{cases}$$

The density metric can be written in matrix format as follows:

$$D = \frac{C^{min}X}{C^{min}O}$$

with $C^{min} = \min(C, C^T)$, $O$ a matrix of identical dimensions as $X$ filled with only 1s, and the division being element-by-element. $C^{min}$ is the symmetrized conditional probability matrix, so that the conditional probability of $p$ on $q$ is equal to that of $q$ on $p$, using the smallest of the two elements ($c_{pq}$ and $c_{qp}$) from $C$. In this way, the asymmetric information in pair-wise conditional probabilities is transformed into a symmetric metric of product 'proximity'.

The $D$ matrix transforms the pair-wise product proximity information into a normalized (in [0,1]) metric of the overall proximity between a given product and all products that are in the current specialization portfolio of a given country. The elements $d_{ij}$ of matrix $D$ are the density of product *i* for country *j* ($D$ has dimensions $m \times n$, like $X$), and they reflect the sum of the cells of the product *j* row of $C^{min}$ for which *i* has a comparative advantage, relative to the sum of all elements in the $j^{\text{th}}$ row of $C^{min}$. Thus, a hypothetical country without any comparative advantages (a country that does not export) would have $d_{ij} = 0$ for all products *j*, and a hypothetical country that has comparative advantage in all products would have $d_{ij} = 1$ for all products *j*.[1]

*2.2. Mutually exclusive RCA: extending density*

The density indicator exploits information about the current (co-)specializations of a country, i.e., which of the entries in the country's column of the $X$ matrix are 1. We propose that there is also valuable information in the patterns in which specializations coincide with absence of specializations, i.e., which of the entries in the country's column of the $X$ matrix are 0. The reason for this is that a country cannot be specialized in all products at the same time, by construction of the RCA indicator.[2]

Thus, if we look at the development pattern of a country, the idea that it will gain specializations as it acquires new capabilities implies also a loss of existing specializations at some point of time. This does not necessarily mean that capabilities are lost, it results from the fact that scarce resources may better be used to exploit other capabilities (in line with basic trade theory). The metaphor of moving through product space implies means that, through competition in global markets, a country can re-position itself into new neighborhoods where some specializations are common and others are (largely) absent.

[1] Note that the computation underlying the density measure is somewhat arbitrary. For example, Coniglio et al. (2021) use a different (and an equally arbitrary) metric where the proximity between a country and a product is computed as the proximity between the given product and the closest one in the specialization portfolio of the country. In a footnote, these authors also indicate that yet another alternative would be to compute the average distance.

[2] This also means that RCA of a country (in a specific product) can change as a result of changes outside the country. For example, if a country $k$ starts exporting more of a particular product *j*, and *ceteris paribus*, $\left({E_{ij}}/{E_j}\right)/\left({E_i}/{E}\right)$ (the non-binary value of RCA) of country *i* will fall, possibly even below 1, so that the binary RCA also changes.

The global pattern of how certain specializations coincide with the absence of some others, or, in other words, that specializations can be mutually exclusive, can reveal valuable information on a different aspect of relatedness. An observation that two specific specializations do not coincide very often may indicate low relatedness, and vice versa. However, the information indicated by the frequent coincidence of one specialization with the absence of another is of a different nature. The latter is an indication of anti-relatedness, thus some sort of competition in specialization.

In order to exploit the potential information in the empirically-observed patterns of such mutual-exclusivity between specializations, we define a complementary conditional probability measure

$$B = U^{-1}ZX^T,$$

where $Z = O - X$ is a matrix of "anti-RCA", in which the elements are $z_{ij} = 1$ if $x_{ij} = 0$ and $z_{ij} = 0$ if $x_{ij} = 1$, and $U$ is the matrix with the row-sums of $Z$ on the main diagonal. Then also the diagonal elements of $U$ $u_{ii} = n - s_{ii}$ are a kind of "anti-ubiquity", i.e., the number of countries that have no RCA in product $j$. The element $b_{pq}$ of matrix $B$ is the probability that a country has RCA in product $q$ conditional on not having RCA in product $p$. These conditional probabilities $b_{pq}$ will capture the "competition" between products for being part of the specialization set of a country that results from trade being based on comparative advantages. Note that the conditional probabilities $c_{pq}$ capture "complementary". Our proposal is that both forms of information are useful for quantifying the product space.

Building on this reasoning, we propose the following indicator:

$$E = \frac{C^T X + B^T Z}{m},$$

which is an $m \times n$ matrix. Clearly, the $C^T X$ term in the numerator bears similarities to the definition of the density measure, although we use the regular $C$ matrix instead of the symmetrized version $C^{min}$. However, matrix $E$ also uses the complementary information that stems from "non-specializations", i.e., the matrix $B^T Z$. Note that for a country $j$ that has actual comparative advantage in $f_j$ products, each entry in the $j^{th}$ column of $C^T X$ is the sum of $f_j$ alternative conditional probabilities, while each entry in the $j^{th}$ column of $B^T Z$ is the sum of $m - f_j$ other conditional probabilities. Thus, through the division by $m$, each entry in the $j^{th}$ column of $E$ qualifies as an average conditional probability (of either the $C$ or $B$ type) over all products.

Matrix $E$ can be seen as a probabilistic re-estimation of $X$ on the basis of the product space. It has the interesting property that each column adds up to the ubiquity of the corresponding product, i.e., $\sum_p e_{pq} = s_{qq}$ (where $e_{pq}$ are the individual elements of $E$). This is why we call $E$ a *ubiquity-redistributing* measure.

The information in matrix $B$ can also be used in a less radical reform of the density measure if we extend the $D$ measure itself with the $B$ matrix. This leads to

$$\widetilde{D} = \frac{C^{min}X + B^{min}Z}{C^{min}O + B^{min}O}$$

where we again use the *min* operator as a way to symmetrize the original $B$ matrix. The numerator of this expression captures the impact of both specialization (RCA = 1) and non-specializations (RCA = 0), whereas in the case of the definition of $D$, it captures only the impact of specializations. The denominator is adjusted accordingly.

### *2.3. Path dependence, related and unrelated diversification*

As discussed, the entries in matrix $E$ are the (ex-post) estimated probabilities for a country having a specialization in a product. We do not want to rule out the possibility that these probabilities are influenced by other factors than product relatedness. For example, the large variety in product ubiquity (the number of countries specialized in the product) suggests that products differ in some inherent way with regard to the extent to how many countries can be specialized in them. The extent of the market or the technological sophistication of the product may be factors that influence this, irrespective of product relatedness.

In order to account for this, we propose a decomposition of the indicator matrix $E$ that is aimed at providing a more precise indication of the impact of product relatedness, and hence is a purer measure of what Coniglio et al. (2021) refer to as path-dependent specializations. For this, we proceed as follows:

$$E = \frac{C^T X + B^T Z}{m} = \frac{C^T X - B^T (O - X)}{m} = \frac{B^T O + K^T X}{m},$$

where $K \equiv C - B$ is a matrix of *marginal* conditional probabilities. Written this way, the matrix $E$ has two (additive) constituent elements. The first of these, $E_1 \equiv B^T O\ /m$ is an $m$ x $n$ matrix in which all rows are equal to each other, i.e., where there is no country-variation. In each of the columns (countries) of this matrix, the $i^{th}$ element indicates an *autonomous* probability of being specialized in product $i$, which we can interpret as the probability for a hypothetical country without any prior comparative advantages (a country that does not trade and begins exporting just product $i$). We characterize this component of matrix $E$ as autonomous because it ignores all information regarding the actual specialization profile of a country. On the other hand, the component $E_2 \equiv K^T X/m$ (also $m$ x $n$) results from the particular specialization profile of the country and can therefore be seen as the path dependent part that corresponds to related diversity.

### *2.4. Product space or country space?*

So far, we formulated probabilities in terms of countries having (or not) a comparative advantage in a specific product. This corresponds to the idea that depicts countries that gain new specializations or loose existing specializations (an element $x_{pq}$ of matrix $X$ turning from 0 to 1 or vice versa) as moving in a product space, which is formed by the network structure of products and their shared capabilities. While this idea is mostly defined at a metaphorical, intuitive and implicit level, it seems most useful in a context where the product space is mostly fixed, or at least changes at a slower time scale than countries move inside the product space.

Our matrix $B$ and its inclusion in the indicators $E$, $E_2$, and $\widetilde{D}$ extends the product space idea by including the notion that product space can only be imagined in terms of specialization, and, hence, that there are also "competitive" relations between products, rather than relations based on only shared capabilities required to produce products.

However, there is also a more radical extension of the idea of product space, which starts from the observation that many countries move very slowly (if at all) in product space. This is evident, for example, from the analysis in Fagerberg and Verspagen (2021), which looks at qualitative and quantitative changes of specializations over the long period 1965 – 2010. Many countries, especially developing countries and even more so, African developing countries remain specialized in a stable set of resource-based products. The developed countries on the other hand show stable specialization patterns in medium and high-tech products. Only a few countries were able to move from resource-based and low-tech specializations into high-tech and medium-tech products.

From this perspective, one could posit the idea of a country space, in which countries are related in a network structure through the productive capabilities that they share. In country space, products "possess" countries that specialize in them, and they may gain (or loose) these specializations. In other words, products move through country space in a similar way that countries move through product space.

The idea of products changing more rapidly may appear strange at first sight, but certainly the nature of products does change over time with the technologies that are used to produce them. One may argue that technology is the fundamental factor that shows a high degree of structural change over time, and that both product space and country space are alternative and abstract notions that reflect different sides of technological change. Hidalgo (2021) mentions the idea of a country space, but does not carry it through to the density measure. We are also not aware of any other published work that extends the concept in order to construct indicators of diversification potential.

In formal terms, all that is required to adopt the idea of country space is to start with a transposed version of matrix $X$: in such a transposed version, the columns correspond to products, and the rows represent their specializations (i.e., countries that specialize in them). In this way, we think of (conditional) probability of product $j$ having country $i$ with comparative advantage. Taking this perspective, we can take the same analytical steps as before, and define country space counterparts of the probability measures defined so far.

We will leave the exact derivations to the interested reader, and simply report these measures below. Note that although we think of a transposed $X$ matrix as the start of the country space idea, we denote all new country space measures in the same $m \times n$ format as used for the product space indicators. We use a * superscript to denote the country space indicators:

$C^* = S^{*-1} X^T X,$

$$D^* = \frac{X^T C^{*min}}{OC^{*min}},$$

$$B^* = U^{*-1} Z^T X,$$

$$K^* \equiv C^* - B^*$$

$$E^* = \frac{XC^* + ZB^*}{n} = \frac{OB^* + XK^*}{n},$$

$$E_1^* \equiv \frac{OB^*}{n} \text{ and } E_2^* \equiv \frac{XK^*}{n},$$

$$\widetilde{D}^* = \frac{XC^{*min} + ZB^{*min}}{OC^{*min} + OB^{*min}}.$$

Now $S^*$ is an $n \times n$ matrix with the column-sums of $X$ on the main diagonal and zeros elsewhere, i.e., the diagonal of $S^*$ contains the countries' diversification scores; $C^*$, $B^*$ and $K^*$are also $n \times n$ matrices, containing conditional probabilities; similarly $U^*$(again $n \times n$) has the count of the lacking comparative advantages of the countries on the diagonal and zeros elsewhere (i.e., $u_{jj}^* = m - s_{jj}^*$). $D^*$, $\widetilde{D}^*$, $E^*$, $E_1^*$ and $E_2^*$ are still $m \times n$ matrices. The columns of $E^*$ add up to the diagonal elements of the matrix $S^*$, or, in other words, the approach with the transposed RCA matrix is *diversity-redistributing*.

The product space and country space indicators can also be aggregated into what we call a "combined space". For matrices $E$ and $E^*$this works as follows. Because $E$ provides probabilities as the average of $m$ conditional probabilities and $E^*$ as the average of $n$ conditional probabilities, the straightforward way to (additively) aggregate the two is

$$E_{Unnorm}^{Tot} = \frac{B^T O + K^T X + OB^* + XK^*}{m+n}.$$

The elements of this matrix add up to the sum of all elements of $X$, which was defined earlier as $r$, but neither its row sums will add up to ubiquity scores nor its column sums to diversity scores. We further normalize this matrix by $r$ into the combined space version of $E$:

$$E^{Tot} = \frac{B^T O + K^T X + OB^* + XK^*}{(m+n)r}.$$

As the elements of $E^{Tot}$ add up to one, the element at $(i, j)$ gives us an estimate based on product-relatedness of what the probability is of finding product *i* and country *j* if we pick a random RCA in the database reflected by $X$. As before, $E^{Tot}$ can be split into an autonomous part and a path dependent part, with $E_1^{Tot} \equiv (B^T O + OB^*)/(m+n)r$ being the autonomous part and $E_2^{Tot} \equiv (K^T X + XK^*)/(m+n)r$ the path-dependent part.

The aggregation of $D$ and $D^*$ yields the combined space versions of the density indicators:

$$D^{Tot} = \frac{C^{min} X + XC^{*min}}{C^{min} O + OC^{*min}},$$

and the aggregation of $\widetilde{D}$ and $\widetilde{D}^*$ leads to

$$\widetilde{D}^{Tot} = \frac{C^{min} X + B^{min} Z + XC^{*min} + ZB^{*min}}{C^{min} O + B^{min} O + OC^{*min} + OB^{*min}}.$$

## 3. Empirical tests

We follow Coniglio et al. (2021) in applying a non-parametric test to investigate the relevance of the ideas that we proposed, and the indicators that embody them. However, rather than the specific test proposed by Coniglio et al, (2021), our test will be based on bootstrapping. We will also follow Coniglio et al (2021) in applying the tests to a subsample of the pooled country-product dataset, focusing both on individual countries, and individual products. Finally, our tests will include tests for the loss of comparative advantage as well as gains.

We have 12 possible indicators (in three groups of four: $D$, $E$, $E_2$, $\widetilde{D}$ are the product space indicators; $D^*$, $E^*$, $E_2^*$, $\widetilde{D}^*$ are the country space indicators; and $D^{Tot}$, $E^{Tot}$, $E_2^{Tot}$, $\widetilde{D}^{Tot}$ are the combined space indicators) that can be used to predict gains and losses of specializations (RCA) in matrix $X$ over time. Rather than applying a regression based that is the most common approach in the literature, we propose a new non-parametric test, and implement it for all 12 indicators.

### *3.1. Data*

We apply the test to data on international trade by countries, for the period 2012 – 2018. The data are taken from the COMTRADE database on the World Bank's WITS server. 155 countries are included, which are all relatively large (over 1 million inhabitants). The value of exports is obtained for 5,197 product categories, which are almost all 6-digit HS (Harmonized System)-2012 categories. Two of the product classes are aggregated groups of 6-digit classes, because for some of the years, some of the 6-digit classes have zero trade among the 155 countries. Export data were downloaded as "mirrored imports" in "all countries" and from the individual 155 countries.

The export value data were converted into binary RCA for each of the years 2012 – 2018. We then construct changes of RCA over all possible combinations of years. Thus, if we look at years $t_1$ and $t_0$, we calculate the matrix $X_1 - X_0$, where the non-zero elements of this matrix are of interest to our analysis: an element equal to 1 indicates an RCA gained, while an element equal to –1 indicates an RCA lost.

The possible changes are grouped by the length of the period. Thus, we have 6 possible 1-year changes (2012-13, 2013-14, etc.), 1 possible 6-year change (2012-18), and all the combinations in between. This yields 21 possible combinations of years, and we implement our test for all 12 indicators with each of those 21 combinations, and for RCAs gained as well as RCAs lost.

### *3.2. Method*

We implement a bootstrapping test for evaluating the predictive power of a specific indicator for either RCA gained or RCA lost. When applied to the entire matrix of RCA changes $(X_1 - X_0)$, this test can be formalized as follows:

$H_0$: $\mathbf{A}_{gain} = \mathbf{A}_{no\text{-}gain}$

where $\mathbf{A}_{gain}$ denotes the average value of the indicator over cells with value 1 in matrix $X_1 - X_0$, and $\mathbf{A}_{no\text{-}gain}$ denotes the average value of the indicator over cells with value not equal to 1 in matrix $X_1 - X_0$. Thus, the null-hypothesis states that the average value of the indicator for cases that (ex post) reveal themselves as actual RCA-gainers is higher than non-gainers.

We construct a $p$-value for the null-hypothesis by the following bootstrapping procedure (Efron and Tibshirani, 1993, chapter 15):

1. Identify the potential RCA-gainers (if we are implementing the test for RCA-gains) or RCA-losers (if we are implementing the test for RCA-losses). The potential RCA-gainers are cells in matrix $X_0$ with value 0. The potential RCA-losers are cells in matrix $X_0$ with value 1. Call the number of potential gainers/losers $N$.
2. Divide the set of potential RCA-gainers or -losers into two subsets: the set of actual gainers or losers, and the set of non-gainers or non-losers. Call the number of actual gainers or losers $N_1$ and note that the number of non-gainers or non-losers is $N - N_1$.
3. Repeat this step 5,000 times: randomly divide the total set of potential gainers or losers (there are $N$ members in this set) into one subset of $N_1$ members and one subset of $N - N_1$ members. Calculate the average of the value in matrix $X_1 - X_0$ of the $N_1$ members in the first subset. Call this average $\mathbf{A}_1$, and compare it to the empirically observed average value $\mathbf{A}_{gain}$.
4. When testing RCA-gains, calculate $p = \#\mathbf{A}_1 \geq \mathbf{A}_{gain} / 5{,}000$, where $\#\mathbf{A}_1 \geq \mathbf{A}_{gain}$ denotes the number of times where $\mathbf{A}_1 \geq \mathbf{A}_{gain}$ holds in step 3. When testing RCA-losses, calculate $p = \#\mathbf{A}_1 \leq \mathbf{A}_{gain} / 5{,}000$, where $\#\mathbf{A}_1 \leq \mathbf{A}_{gain}$ denotes the number of times where $\mathbf{A}_1 \leq \mathbf{A}_{gain}$ holds in step 3. $p$ is the observed $p$-value for $H_0$.

Note that the number of 5,000 repetitions in step 3 of the procedure is arbitrary (Efron and Tibshirani, 1993 use 1,000 repetitions in their examples, so our 5,000 repetitions are on the safe side).

For a specific combination of years and an indicator, we apply this procedure in three different ways, which differ with regard to the sample to which the test if applied. The most straightforward way is to apply the test the entire matrix $X_1 - X_0$, which implies that there are 155 × 5,197 = 805,535 observations that enter the procedure at step 1. The other two ways are to run the test either by product (individual rows of the matrix $X_1 - X_0$; 155 observations enter at step 1) or by country (individual columns of matrix $X_1 - X_0$; 5,197 observations enter at step 1).

Note also that small values of $N_1$ will lead to weak test results: imagine the extreme case where $N = N_1 = 1$; this prevents any variety in step 3 of the procedure and hence reduces the test to the question of whether the single value of the observation with RCA-gain or -loss is above or below (respectively) the non-gainers or non-losers. With $N > 1$ but still "small", we have a less extreme case, but still a relatively weak test. This happens especially when we consider RCA-loss, because the group of potential RCA-losers (observations that have RCA) is usually (much) smaller than the case of potential gainers. The case of small $N$ is also limited to cases where we implement tests on either individual countries or products. In the empirical implementation of the tests below, we limit the cases that we subject to the test to those where $N > 15$ (admittedly an arbitrary cut-off).

### *3.3. Data description*

On a yearly basis, there are almost 82 thousand RCAs (cells of matrix $X$ equal to 1), which is about 10% of the total (Table 1). The standard deviation of the number of RCAs is small (591). There are three years with a slightly above-average number, and four years with below-average values. The first year, 2012, has the lowest number of RCAs.

**Table 1. Number of RCA in matrix *X*, yearly**

| Year | Number of RCAs (elements = 1 in matrix *X*) | Fraction of total |
|---|---|---|
| 2012 | 80,540 | 0.100 |
| 2013 | 82,338 | 0.102 |
| 2014 | 82,268 | 0.102 |
| 2015 | 81,252 | 0.101 |
| 2016 | 81,561 | 0.101 |
| 2017 | 82,083 | 0.102 |
| 2018 | 81,773 | 0.102 |
| Average | 81,688 | 0.101 |

For RCA changes, we use the simple subtraction of binary RCA, which is a rough measure because it does not distinguish between small and large changes. For example, a change from 0 to 1 (a gain of specialization) could occur because $\left(E_{ij}/E_j\right)/\left(E_i/E\right)$ changes from 0.99 to 1.01 (a marginal change), or because it changes from 0 to 1.5 (a large change).[3] The analysis in Appendix 1 suggests that small changes do not dominate the results, but also that, as could be expected, small changes are relatively more frequent for shorter periods.

Table 2 documents the descriptive statistics on the number of RCA changes, which is the main variable in our analysis. The table is organized by length of the periods over which changes are considered, and documents all 21 possible combinations of years. We generally see an increase of the number of changes with the length of the period, with the number of gains fluctuating between 13,360 (2014-15) and 25,300 (2012-18) and the number of losses fluctuating between 13,410 (2012-13) and 24,067 (2012-18).

The table also documents the averages of the number of changes per country, and per product. Here again, the long 6-year period shows the largest numbers, with an average 163 gains per country (out of 5,197 products) and almost 5 (out of 155 countries) average gains per product. There are 155 RCA losses on average per country in this long period, and slightly over 4½ losses per product. The standard deviations of all averages in the table are relatively large, although always smaller than the averages themselves.

[3] Coniglio et al. (2021) require a larger jump for a gain of RCA to have taken place.

**Table 2. Descriptive statistics on RCA changes**

| | # gains | Av # gains per country (stdev) | Av # losses per country (stdev) | # losses | Av # gains per product (stdev) | Av # losses per product (stdev) |
|---|---|---|---|---|---|---|
| One-year changes | | | | | | |
| 2012-13 | 15,208 | 98.1 (71.1) | 86.5 (67.3) | 13,410 | 2.93 (2.49) | 2.58 (2.19) |
| 2013-14 | 13,748 | 88.7 (66.0) | 89.1 (65.9) | 13,818 | 2.65 (2.28) | 2.66 (2.23) |
| 2014-15 | 13,360 | 86.2 (56.3) | 92.7 (71.9) | 14,376 | 2.57 (2.26) | 2.77 (2.29) |
| 2015-16 | 13,765 | 88.8 (56.5) | 86.8 (63.3) | 13,456 | 2.65 (2.36) | 2.59 (2.26) |
| 2016-17 | 13,598 | 91.1 (63.9) | 87.7 (62.3) | 13,598 | 2.72 (2.30) | 2.62 (2.39) |
| 2017-18 | 13,826 | 89.2 (71.3) | 91.2 (61.8) | 14,136 | 2.66 (2.34) | 2.72 (2.36) |
| Two-year changes | | | | | | |
| 2012-14 | 18,882 | 121.8 (90.3) | 110.7 (87.1) | 17,154 | 3.63 (2.85) | 3.30 (2.53) |
| 2013-15 | 16,779 | 108.3 (70.6) | 115.3 (89.1) | 17,865 | 2.23 (2.60) | 3.44 (2.61) |
| 2014-16 | 17,242 | 111.2 (68.9) | 115.8 (90.9) | 17,949 | 3.32 (2.70) | 3.45 (2.62) |
| 2015-17 | 17,686 | 114.1 (78.7) | 108.7 (82.3) | 16,855 | 3.40 (2.67) | 3.24 (2.66) |
| 2016-18 | 17,666 | 114.0 (83.8) | 112.6 (70.9) | 17,454 | 3.40 (2.67) | 3.36 (2.77) |
| Three-year changes | | | | | | |
| 2012-15 | 20,490 | 132.2 (87.1) | 127.6 (103.0) | 19,778 | 3.94 (3.08) | 3.81 (2.75) |
| 2013-16 | 19,558 | 126.2 (80.2) | 131.2 (105.4) | 20,335 | 3.76 (2.94) | 3.91 (2.81) |
| 2014-17 | 19,693 | 127.1 (84.3) | 128.2 (101.9) | 19,878 | 3.79 (2.88) | 3.82 (2.85) |
| 2015-18 | 19,859 | 128.1 (93.3) | 124.8 (85.7) | 19,338 | 3.82 (2.85) | 3.72 (2.88) |
| Four-year changes | | | | | | |
| 2012-16 | 22,725 | 146.6 (93.6) | 140.0 (116.7) | 21,704 | 4.37 (3.34) | 4.18 (2.90) |
| 2013-17 | 21,604 | 139.4 (94.6) | 141.0 (115.9) | 21,859 | 4.16 (3.08) | 4.21 (2.99) |
| 2014-18 | 21,295 | 137.4 (97.4) | 140.6 (105.9) | 21,790 | 4.10 (2.99) | 4.19 (3.00) |
| Five-year changes | | | | | | |
| 2012-17 | 24,451 | 157.7 (107.5) | 147.8 (129.1) | 22,908 | 4.70 (3.46) | 4.41 (3.02) |
| 2013-18 | 22,777 | 146.9 (104.9) | 150.6 (114.5) | 23,342 | 4.38 (3.12) | 4.49 (3.11) |
| Six-year change | | | | | | |
| 2012-18 | 25,300 | 163.2 (115.8) | 155.3 (124.7) | 24,067 | 4.87 (3.44) | 4.63 (3.12) |

### *3.4. Offsetting product ubiquity in matrix $E_2$ and $E_2^*$*

We now illustrate the working of the decomposition of matrix $E$ or $E_2^*$ into the autonomous part $E_1$ or $E_1^*$ on the one hand and the path dependent path $E_2$ or $E_2^*$ on the other hand. We show results based on the comparative advantage matrix ($X$) for 2012, but other years show similar results. The top part of Figure 1 shows the relation between the autonomous probability (elements of matrix $E_1$ or $E_1^*$) on the horizontal axis and ubiquity on the vertical axis. This relationship is very tight, and the slopes of the regression lines are very similar to the number of countries (in product space) or the number of products (in country space), which suggests that dividing ubiquity by the number of countries is a good approximation for the autonomous probabilities in $E_1$, and dividing by ubiquity (of countries) is a good approximation for the autonomous probabilities in $E_1^*$.

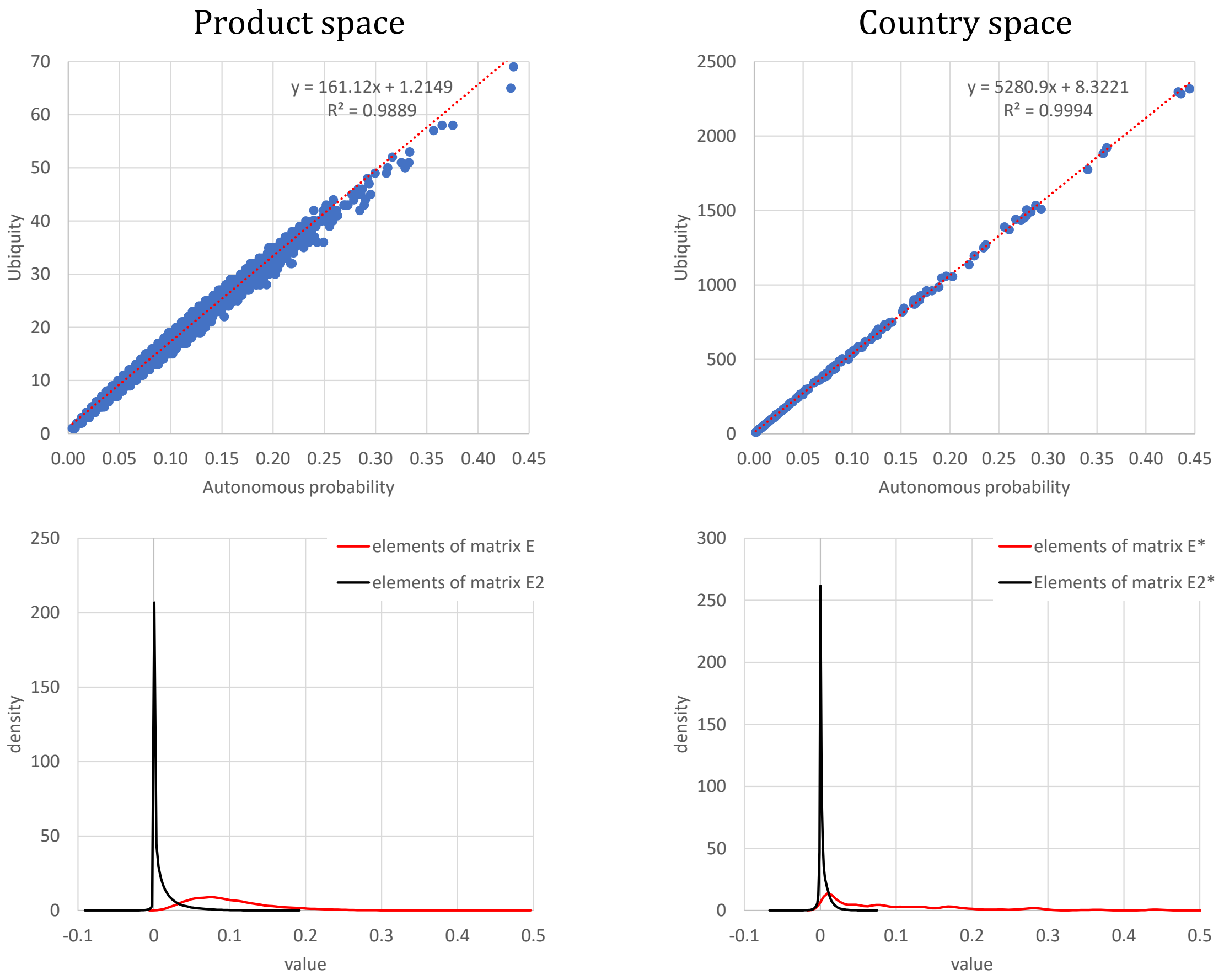

**Figure 1. The impact of the decomposition of matrices $E$ or $E^*$**

The bottom part of Figure 1 shows kernel density plots for the entire set of elements of the matrices $E$ and $E_2$ (product space on the left side of the figure) or $E^*$ and $E_2^*$ (country space on the righthand side). This shows that these matrices are very markedly different. The matrices with pure path dependence indicators ($E_2$ and $E_2^*$) show a very peaked distribution as compared to the matrices $E$ and $E^*$, and the peak of their distributions lies much closer to the origin. This implies that the pure path dependency measures show smaller differences between either countries (product space) or products (country space), or, that the matrices $E$ and $E^*$ (or any of the D-type matrices) may exaggerate the impact of path dependence on specialization patterns.

### *3.5. Statistical results*

We now proceed to test the predictive power of the indicators, both in terms of predicting gains of specializations, and losses. The story for tests on the entire matrix $X_1 - X_0$ is rather simple: all these 504 tests ( 21 combinations of years × 12 indicators × 2 for gains and losses) show $p$-values < 0.0001. This means that at the level of the entire dataset (which is what most empirical tests so far have been aimed at), all indicators that we proposed are relevant for predicting RCA gains as well as losses.

Is this any different if we implement the tests on separate rows (products) or columns (countries) of the matrix $X$? In order to investigate this, we constructed diagrams that show the cumulative distribution of the $p$-values of the tests. The $p$-values are displayed on the (logarithmically scaled) horizontal axis, the fraction of all observations with $p$-value equal to or smaller is on the vertical axis. In order to save space, we pool al possible time intervals (from 1-year to 6-year intervals) here in the main text. We report summary results for individual period lengths in Appendix 2. The pooling does not change the relative predictive performance of the indicators much, although the results for separate time intervals show that predictive power generally gets higher when the length of the interval increases. This could well be related to our finding (in Appendix 1) that small changes in RCA (which could be considered as statistical noise) are more frequent for shorter periods.

Figure 2 shows the tests for individual products. This asks how well, for a particular individual product, the indicators can predict the gain or loss of RCA in the product by the 155 countries in the database. The separate panels group the individual indicators into product space type, country space type and combined space type, and present tests for gain of RCA and loss of RCA separately. Note that in the top row (product space indicators), $E$ and $E_2$ yield identical results by construction of these indicators. In the bottom row (combined space), these indicators yield different results, but these differences are so small that they are undistinguishable in the figures.

Several things stand out from the comparison between those categories. First, we see that at the product level, the power of the tests is much lower than at the level of the entire matrix. The fraction of products that shows $p$-value ≤ 0.01 is never above 30% (roughly 1,500 out of 5,197 products). At $p \leq 0.10$, the maximum share of products is about 60%, but is usually lower.

Second, we see that without exception, the proposed indicators perform worse for loss of RCA than for gain of RCA, or, in other words, RCA losses are harder to predict than RCA gains. The lines in the panels on the righthand side (loss of RCA) are always lower than on the lefthand side (gain of RCA). With $p \leq 0.01$, only about 10% of products shows significant results, and with $p \leq 0.10$, a maximum of about 45% of products pass the test.

Third, we see that the difference between the various indicators is usually small, except within the group of country space indicators, where variety between indicators is relatively large for both gains and losses, and where the density indicator $D^*$ clearly performs best for RCA gains. $D^*$ is the best performing indicator (in the entire range left of $p = 0.10$) for RCA gains. For RCA losses, differences between indicators are even smaller. In this case, $\widetilde{D}^{Tot}$ is marginally on top as the best performing indicator.

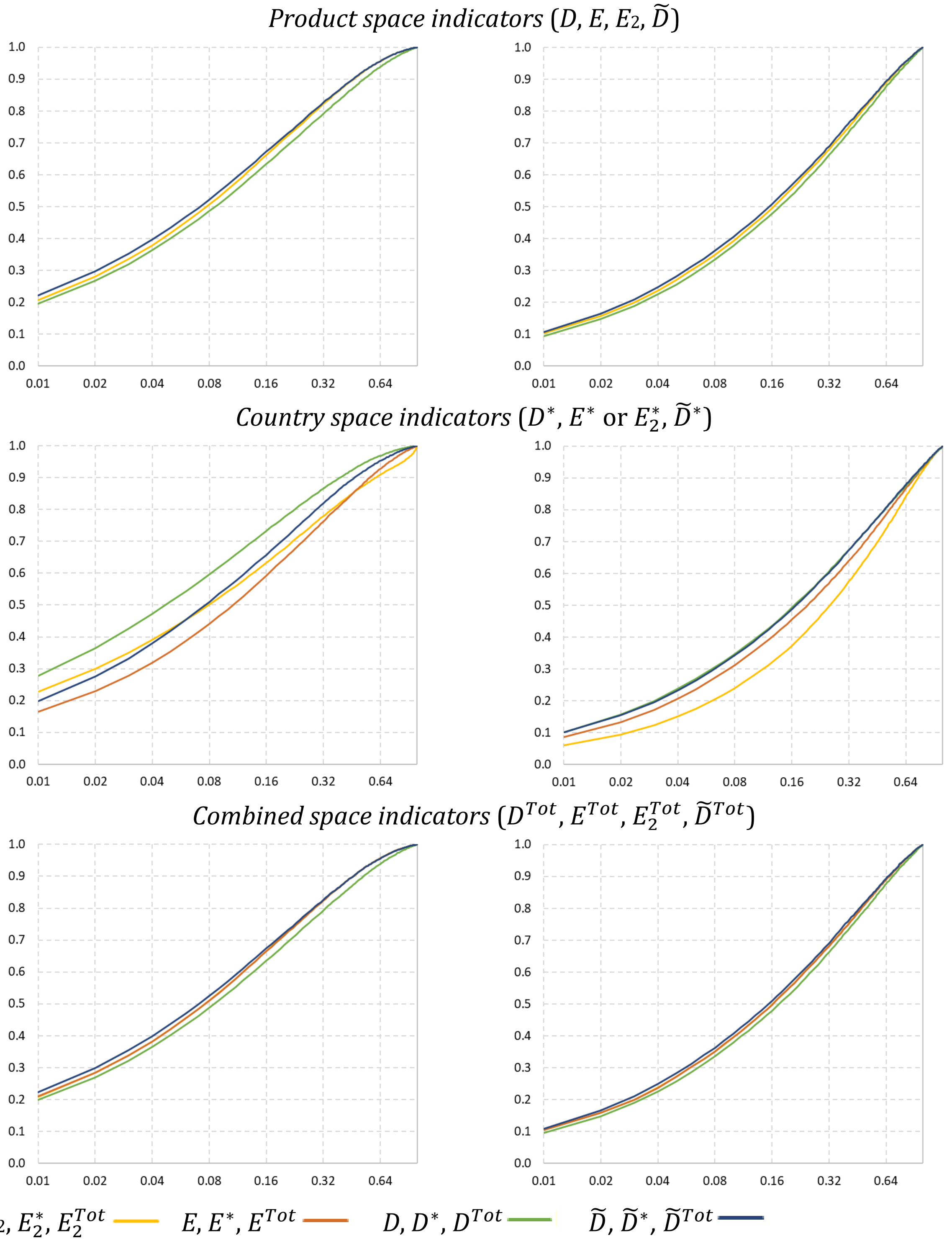

**Figure 2. Tests on individual products (rows of matrix $X$), gains of RCA (lefthand side) and loss of RCA (righthand side)**

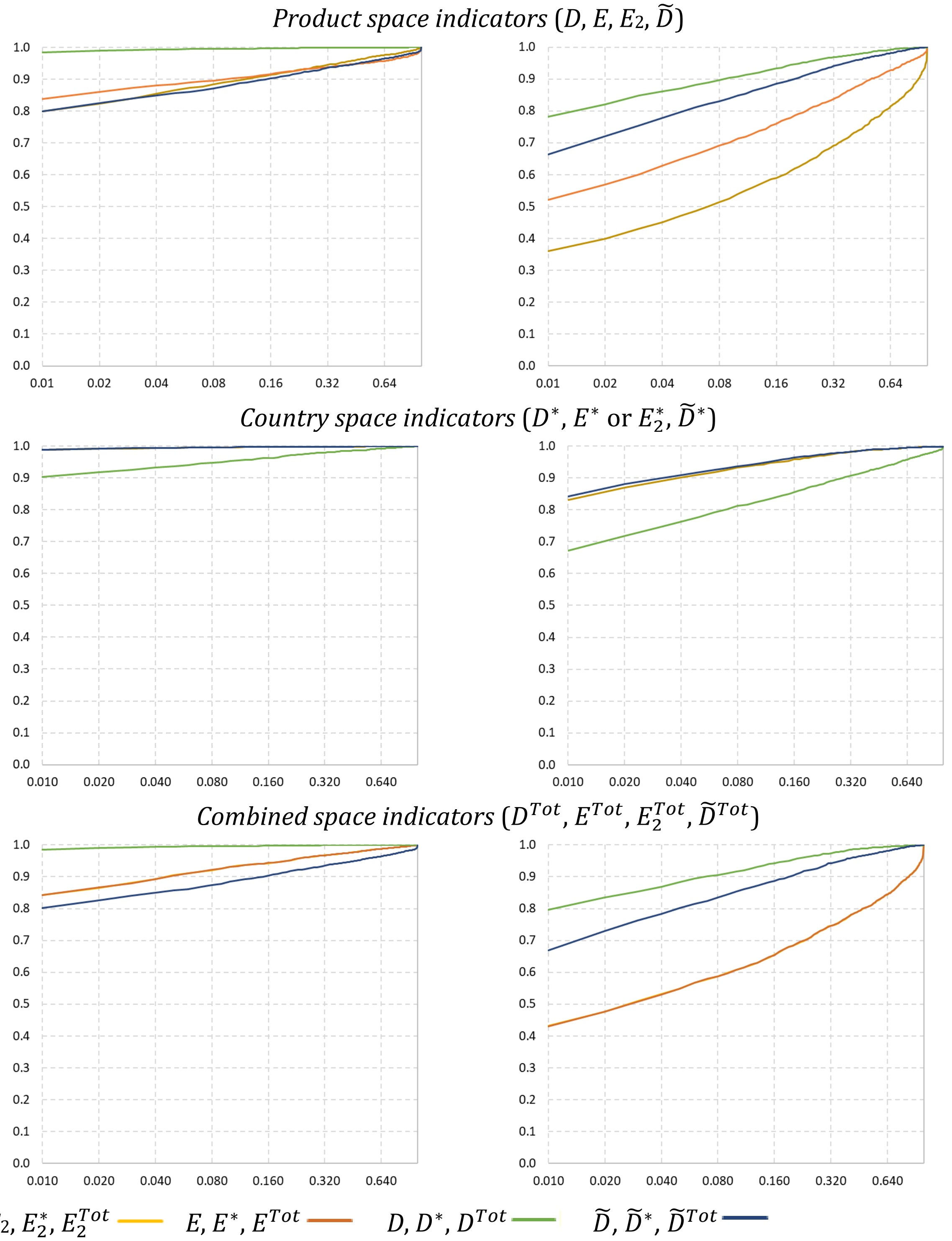

**Figure 3. Tests on individual countries (columns of matrix *X*), gains of RCA (lefthand side) and loss of RCA (righthand side)**

Figure 3 shows the results for tests on individual countries, arranged in the same format. Here $E^*$ and $E_2^*$ are equivalent, and the pair $E^{Tot}$ and $E_2^{Tot}$ yields results that are slightly different, but not distinguishable in the graphs. These tests are more powerful than the tests for products, i.e., it seems easier to predict which products a country will gain or lose RCA in than it is to predict which countries will gain or lose RCA in a specific product.

**Table 3. Summary of results in Figure 2 and 3 for cutoff $p$-values 0.01, 0.05 and 0.1**

| | p ≤ 0.01 | p ≤ 0.05 | p ≤ 0.10 | p ≤ 0.01 | p ≤ 0.05 | p ≤ 0.10 |
|---|---|---|---|---|---|---|
| | Gains - Products | | | Losses - Products | | |
| | Product space | | | Product space | | |
| $E$ and $E_2$ | 0.207 | 0.418 | 0.554 | 0.104 | 0.270 | 0.393 |
| $D$ | 0.197 | 0.401 | 0.531 | 0.095 | 0.255 | 0.377 |
| $\widetilde{D}$ | **0.221** | **0.434** | **0.568** | **0.108** | **0.282** | **0.405** |
| | Country space | | | Country space | | |
| $E_2^*$ | 0.227 | 0.424 | 0.542 | 0.060 | 0.175 | 0.278 |
| $E^*$ | 0.165 | 0.355 | 0.486 | 0.085 | 0.237 | 0.355 |
| $D^*$ | <u>0.278</u> | <u>0.510</u> | <u>0.639</u> | **0.102** | **0.270** | **0.389** |
| $\widetilde{D}^*$ | 0.198 | 0.418 | 0.556 | 0.101 | 0.264 | 0.384 |
| | Combined space | | | Combined space | | |
| $E_2^{Tot}$ | 0.209 | 0.421 | 0.557 | 0.105 | 0.271 | 0.395 |
| $E^{Tot}$ | 0.210 | 0.422 | 0.557 | 0.104 | 0.272 | 0.395 |
| $D^{Tot}$ | 0.198 | 0.402 | 0.533 | 0.096 | 0.257 | 0.379 |
| $\widetilde{D}^{Tot}$ | **0.224** | **0.437** | **0.571** | <u>0.109</u> | <u>0.283</u> | <u>0.407</u> |
| | Gains - Countries | | | Losses - Countries | | |
| | Product space | | | Product space | | |
| $E_2$ | 0.798 | 0.865 | 0.894 | 0.361 | 0.472 | 0.540 |
| $E$ | 0.839 | 0.885 | 0.902 | 0.521 | 0.648 | 0.713 |
| $D$ | **0.985** | **0.994** | <u>0.996</u> | **0.782** | **0.871** | **0.908** |
| $\widetilde{D}$ | 0.799 | 0.856 | 0.883 | 0.664 | 0.797 | 0.848 |
| | Country space | | | Country space | | |
| $E_2^*$ and $E^*$ | 0.987 | **0.994** | 0.996 | 0.831 | 0.910 | 0.941 |
| $D^*$ | 0.902 | 0.937 | 0.952 | 0.671 | 0.778 | 0.824 |
| $\widetilde{D}^*$ | <u>0.988</u> | **0.994** | <u>0.996</u> | <u>0.843</u> | <u>0.917</u> | <u>0.943</u> |
| | Combined space | | | Combined space | | |
| $E_2^{Tot}$ | 0.842 | 0.905 | 0.932 | 0.432 | 0.548 | 0.607 |
| $E^{Tot}$ | 0.843 | 0.903 | 0.931 | 0.431 | 0.549 | 0.608 |
| $D^{Tot}$ | **0.984** | <u>0.995</u> | <u>0.996</u> | **0.796** | **0.883** | **0.916** |
| $\widetilde{D}^{Tot}$ | 0.802 | 0.857 | 0.886 | 0.669 | 0.803 | 0.853 |

Notes: values that are underlined are the maximum in their category (gain – products; losses – products; gains – countries; losses – countries), values that are printed in bold are the maximum in their subcategory (gain – products; losses – products; gains – countries; losses – countries *and* product space; country space; combined space).

$D$, $E^*/\ E_2^*$, $\widetilde{D}^*$ and $D^{Tot}$ predict RCA gains almost perfectly (99% or more of cases has $p \leq 0.01$), while for RCA losses, $E^*/\ E_2^*$ and $\widetilde{D}^*$ (both are in the country space group) are the best performing indicators. Variety in performance is generally larger for the country-based tests in Figure 3 than the product-based tests in Figure 2.

Table 3 provides a summary of the results in Figures 2 and 3 for the cutoff $p$-values 0.01, 0.05 and 0.1. This brings out the relatively strong performance of the country space indicators. For predicting gains of specializations by product, the country space indicator $D^*$ attains the highest fraction cases at all three cutoff $p$-values. For predicting gains of specializations by country, the country space indicator $\widetilde{D}^*$ is on top for two out of three $p$-values (0.01 and 0.1), and for predicting losses of specializations by country, the country space indicator $\widetilde{D}^*$ is on top for all three $p$-value cutoffs. The country space idea seems perhaps to be more relevant than the product space idea in our data.

We now turn to looking at subgroups of products. For this, we use the so-called Lall concordance, which divides the 5,197 products in our database into 11 groups (Lall, 2000).[4] In order to economize on space, we document only a part of the entire cumulative distributions that were displayed in the previous figures. These results are presented in Figures 4 – 6, with each figure presenting a "different space". In these figures, the vertical axis shows the fraction of observations with $p \leq 0.10$, split into three groups (with cut-off points 0.05 and 0.01).

These results confirm some of the impressions from the previous figures, such as the relative difficulty of predicting loss of RCA, which is present throughout all 11 Lall categories, and the fact that in many cases there are only small differences between the indicators. However, there are also interesting differences between the Lall categories.

Generally, primary products, resource-based products and low-tech products (i.e., the first six groups) are harder to predict (lower bars in the graphs) than the five medium- and high-tech groups. The highest bar is observed for $D^*$ predicting RCA gains in electronics and electrics (high-tech), while the lowest bar for gains is found for $E^*$ in primary agricultural products. We also observe that differences between the indicators seem to be somewhat less pronounced in the product space group (Figure 4), relative to the country space and combined space groups (Figures 5 and 6).

There are also some interesting differences between the product space, country space and combined space indicators. For example, while within the country space group, $D^*$ performs well relative to other indicators in predicting RCA gains, the corresponding indicators $D$ (product space group) and $D^{Tot}$ (combined space group) generally perform somewhat less well.

[4] The original Lall concordance has 10 groups, but we split the primary products group into two, one for mineral resources, and the other agro-related.

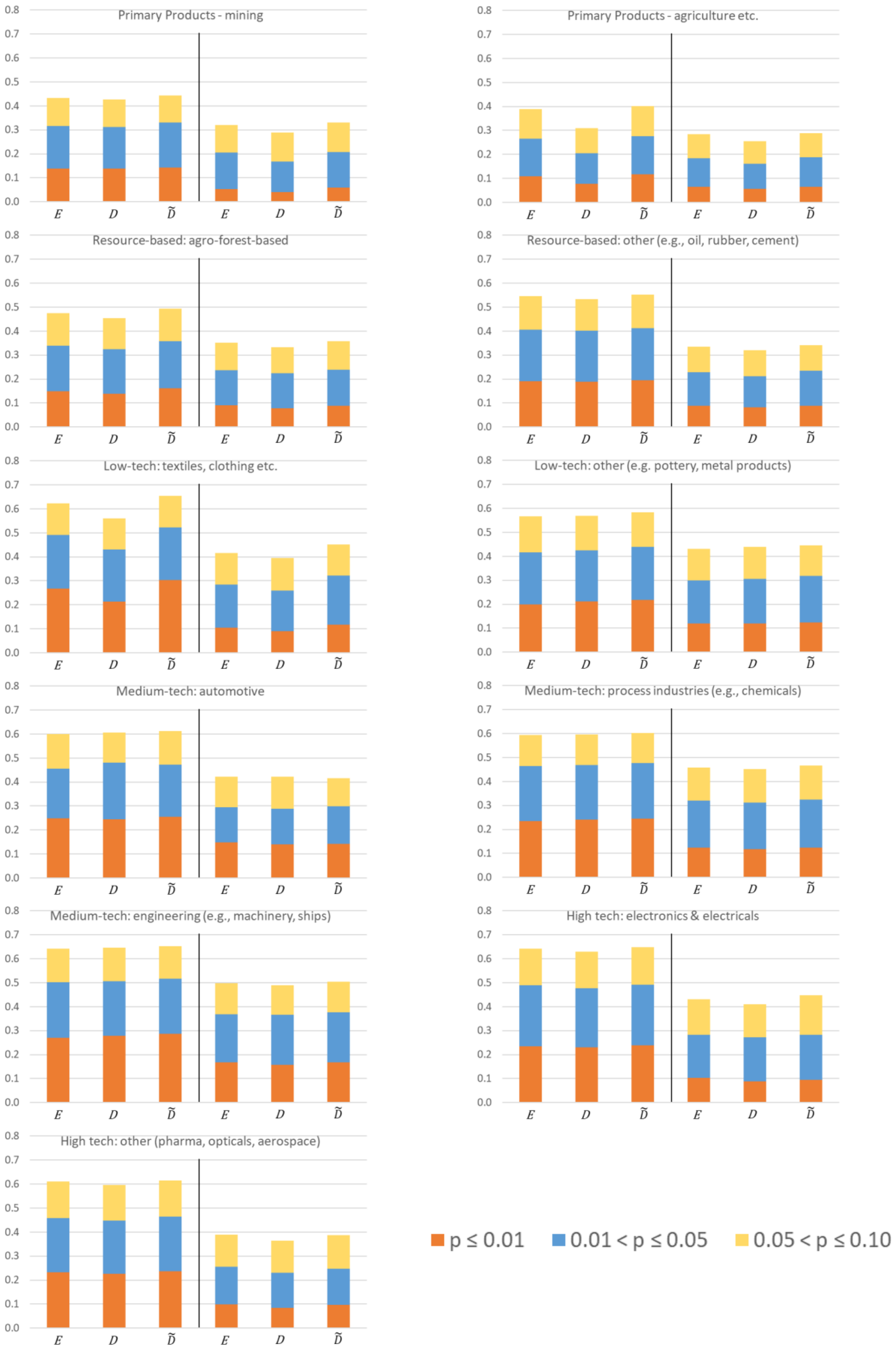

**Figure 4. Product space indicators for separate Lall groups** (vertical axis of each graph contains fractions of observations at indicate significance level; each graph contains results for RCA gains, left of the vertical line, and loss of RCA, right of vertical line)

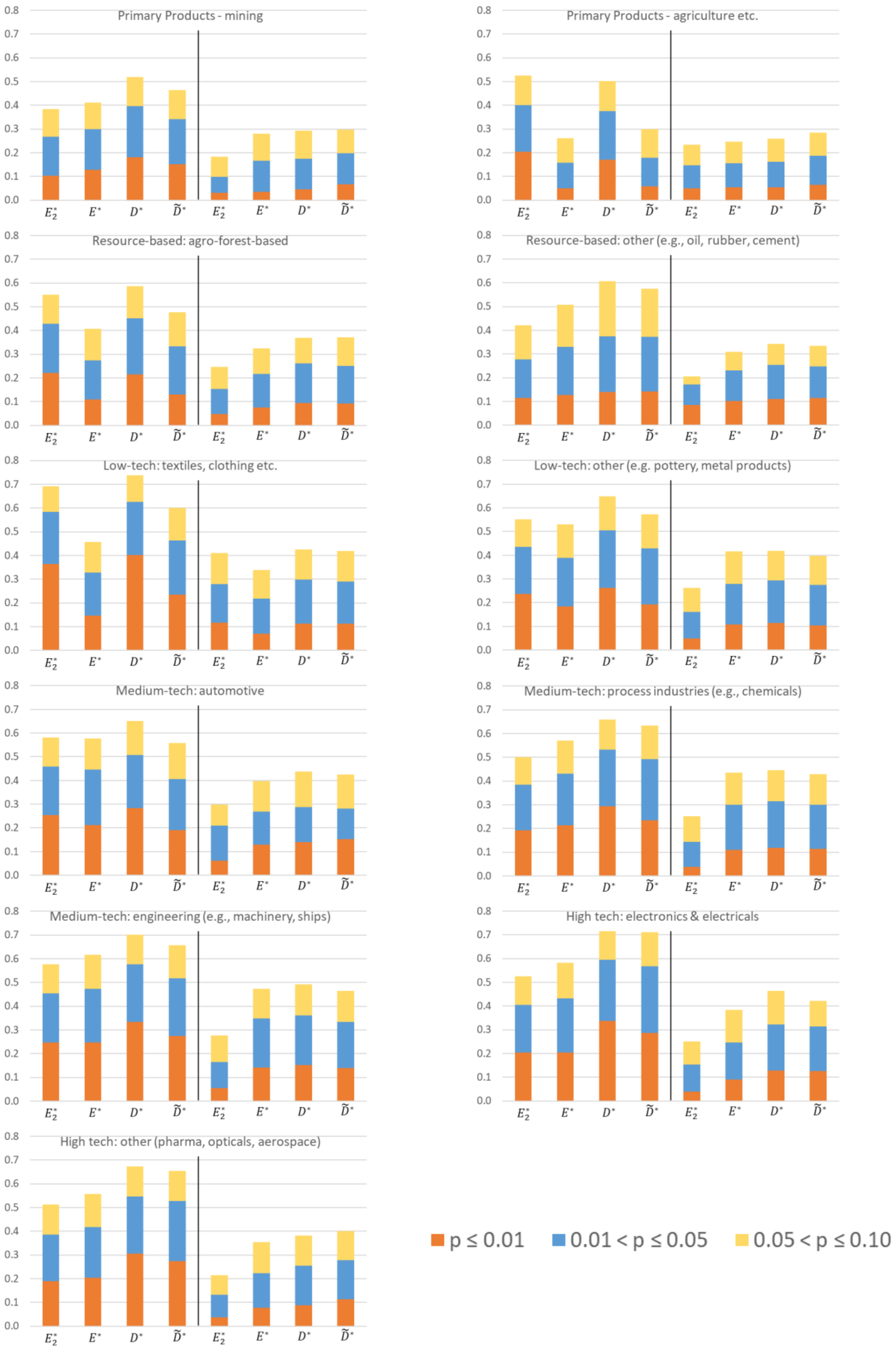

**Figure 5. Country space indicators for separate Lall groups** (vertical axis of each graph contains fractions of observations at indicate significance level; each graph contains results for RCA gains, left of the vertical line, and loss of RCA, right of vertical line)

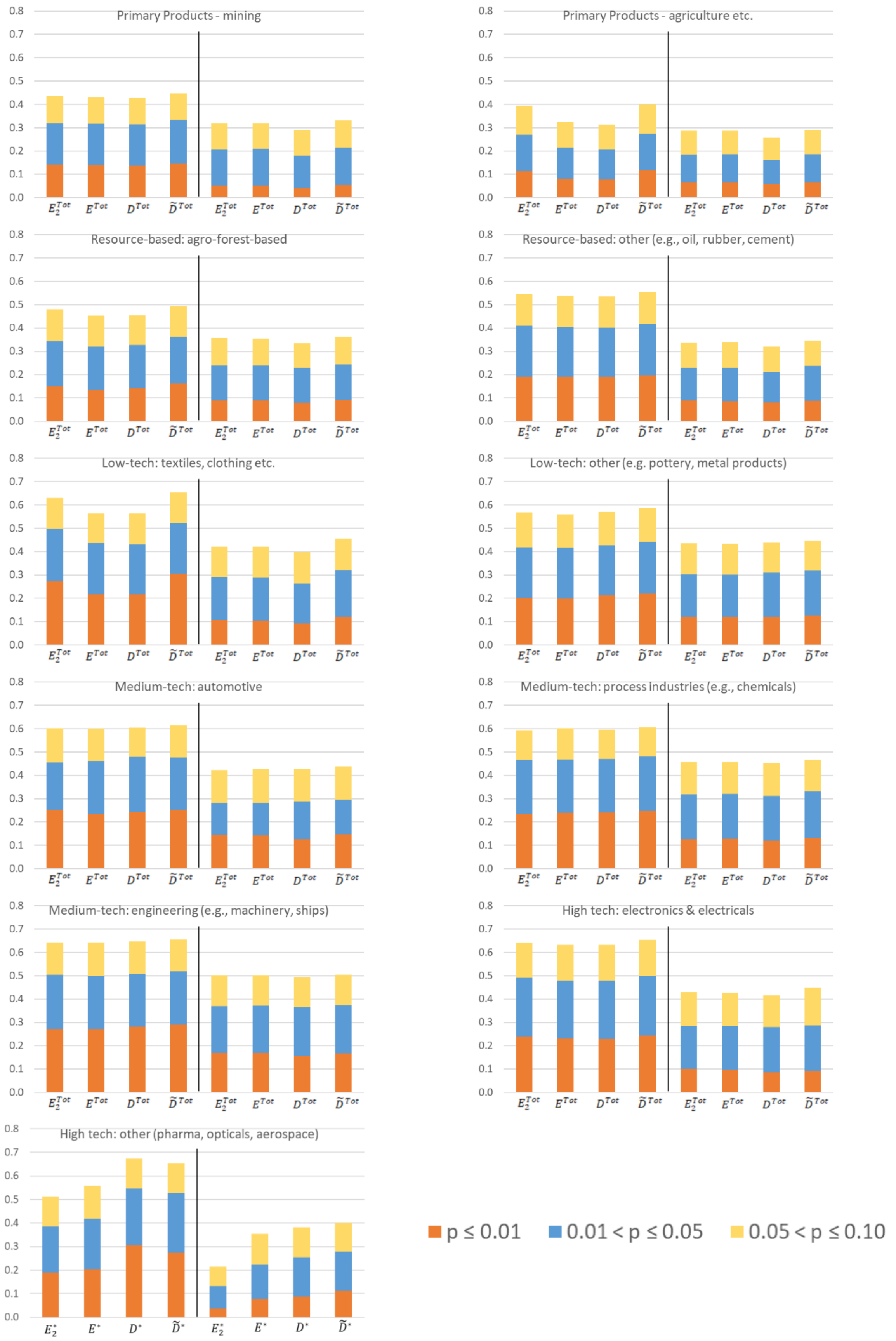

**Figure 6. Combined space indicators for separate Lall groups** (vertical axis of each graph contains fractions of observations at indicate significance level; each graph contains results for RCA gains, left of the vertical line, and loss of RCA, right of vertical line)

### *3.6. Summary of results*

In terms of theoretical contributions, our four proposals had two main messages. The first is that absence or loss of comparative advantage should be considered as a useful source of information. The second is the idea of country space, or a combined product and country space. The empirical analysis reported on above suggests that these ideas contribute relatively most when we look either at individual products or individual countries. When analyzing a pooled country and products sample, all measures that we analyzed performed extremely well, and hence the theoretical proposals do not add much to an already high predictive power.

In terms of the absence of comparative advantage, the indicators $D$, $D^*$ and $D^{Tot}$ do not include this, while all other indicators include this type of information. Does this lead to any differences in performance? This greatly depends on the specific context in which the indicators are used. In product space, i.e., indicator $D$ and the other non-starred or non-*Tot* indicators, $D$ does better in predicting gains or losses of RCA per country, but it does slightly worse for predicting gains or losses of RCA per product.

In terms of predicting losses of comparative advantage, which in our view, needs to be a more integral part of the literature, the results are very clear: it is much harder to predict loss of comparative advantage than to predict gains. This holds for all indicators across the board.

In terms of our second theoretical proposal, product space, in our analysis the country space indicators generally come out with stronger predictions than the product space indicators, suggesting that country space is a very relevant notion. Developing the country space idea further, both theoretically and empirically, could therefore be a useful avenue for the literature.

## 4. Summary and conclusions

The product space literature provides an interesting and fruitful perspective on the old questions of how countries can grow and develop to increase the living standards of their inhabitants in a quantitative and qualitative way. In this paper, we proposed a number of additions to this idea, which add a number of new intuitions and according indicators to the already existing discourse in this literature. Our empirical tests, although performed on a specific dataset covering a short time period, show that these proposals have empirical relevance.

In particular, we argued that the absence and loss of specializations should become a part of the literature, just as the presence and gain of specializations already is. This suggestion stems from the fundamental premises of trade theory (i.e., that countries deploy their resources to activities in which they have *comparative* advantage), and we believe that application of this idea in the product space literature would make it possible to integrate this field better with trade theory. This could have important advantages for how the product space idea develops further in the field of economic theory at large.

We also argued that the product space idea has an intuitive counterpart in the notion of country space. Especially this idea appears to be empirically relevant. With product space, we think of countries moving through a (relatively) fixed landscape made up of products and their relatedness. Country space, on the other hand, describes products as moving through a (relatively) fixed landscape of countries. We believe that the elaboration of this idea would enable the product space literature to develop a more explicit and more coherent view on the structural mechanisms that enable countries to develop. It would also provide a way of better integrating with the large and heterogenous literature on development and technological change (e.g., Freeman and Perez, 1988; Grossman and Helpman, 1991; Verspagen, 1991; Fagerberg, 1994; Von Tunzelmann, 1995).

## Appendix 1. RCA changes

In order to assess the magnitude of changes in RCA (either gains or losses), we created a binary measure of RCA, as follows:

$$\chi_{ij} = \frac{E_{ij}/E_j}{E_i/E} \text{ and } \rho_{ij} = \frac{\chi_{ij}-1}{\chi_{ij}+1}$$

where, as in the main text, $E_{ij}$ denotes the value of exports of product *i* by country *j*, and the absence of a subscript indicates summation over the relevant dimension. Obviously, $\chi_{ij}$ is a continuous (on the interval $[0,\infty\rangle$) rather than a binary measure. The transformation of $\chi_{ij}$ to $\rho_{ij}$ makes the measure symmetric, in the interval $[-1,1]$. Thus, the absolute value of the largest possible change of $\rho_{ij}$ from one period to another is 2.

For all periods as specified in the main text, we first created the changes of binary RCA, either gains (from 0 to 1), or losses (from 1 to 0). For each of the two categories of gains and losses, we then estimated the kernel density (i.e., estimated probability density function) of the distribution of changes in $\rho_{ij}$. These results are documented in the figure below, for different period lengths. We pool all possible combinations of years for each period length, which implies that shorter period lengths have more observations than longer period lengths.

There are two main features that emerge from the figure. First, we see that both losses and gains of binary RCA result from a wide range of magnitude of the underlying continuous changes. Second, we notice that the longer the time period gets, the flatter the distribution becomes. The distribution for 1-period changes is fairly peaked on the left side, but this gradually becomes much less prominent when we move towards the 6-year period.

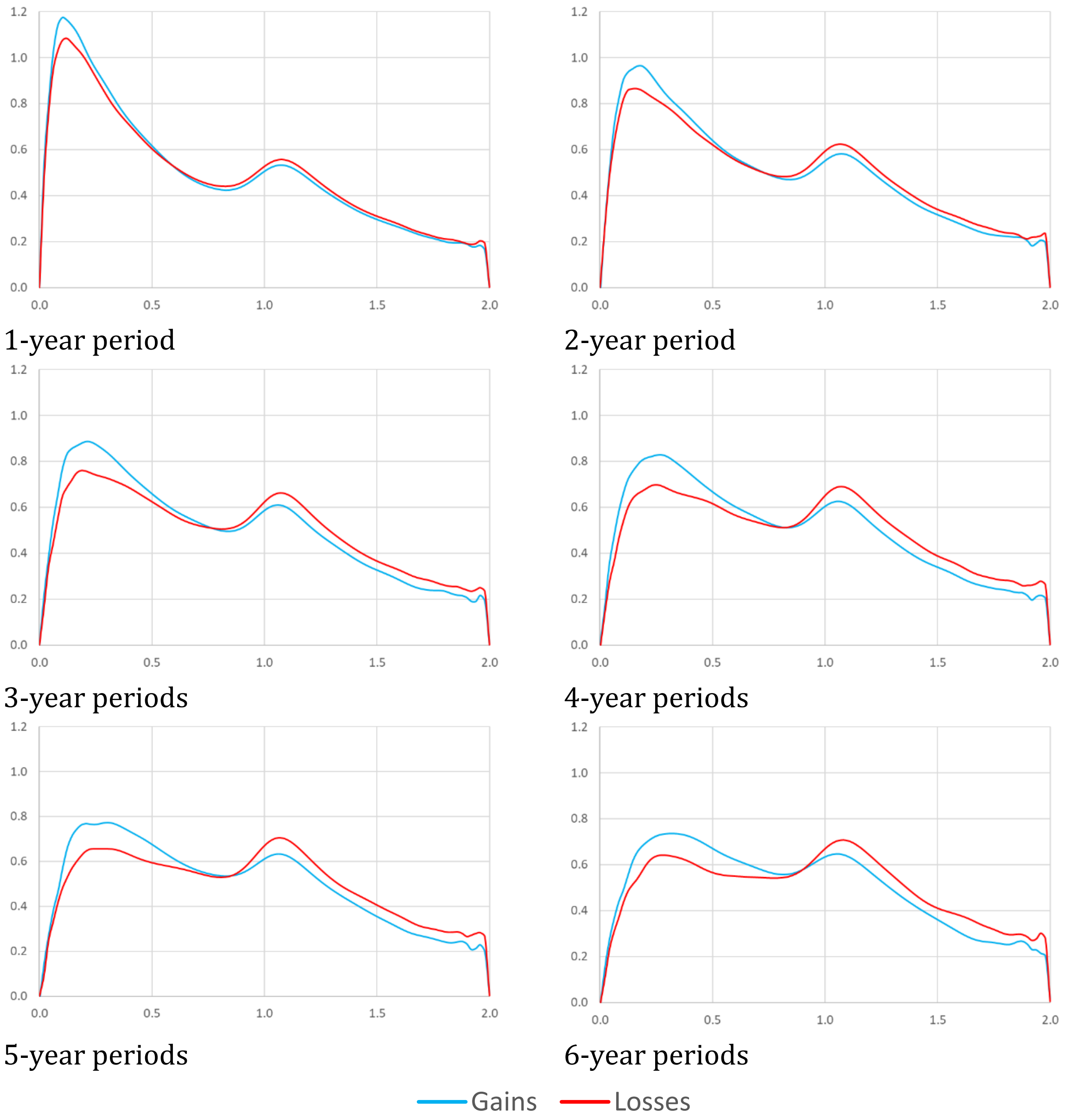

**Figure A1. Kernel density plots of magnitude of RCA change, gains of RCA and losses of RCA**

## Appendix 2. Additional bootstrapping test results

In this appendix we document bootstrapping test results for individual period lengths. The main text explains the setup and execution of the tests, and documents the results for the pooled period lengths, i.e., all period lengths from 1 period to 6 periods. The tables in this appendix show that there are some differences between the subsets according to period length, and between those subsets and the pooled set. The estimates for individual period lengths indicate that predictive power generally gets higher when the length of the interval increases. However, the general conclusions are common and well represented in the pooled set.

As in the main text, values that are underlined are the maximum in their category (gain – products; losses – products; gains – countries; losses – countries), and values that are printed in bold are the maximum in their subcategory (gain – products; losses – products; gains – countries; losses – countries and product space; country space; combined space). Whenever underlining or bold face changes relative to Table 3 in the main text, the values in the appendix tables below are printed in red.

**Table A1. Summary of test results for 1-period changes**

| | p ≤ 0.01 | p ≤ 0.05 | p ≤ 0.10 | p ≤ 0.01 | p ≤ 0.05 | p ≤ 0.10 |
|---|---|---|---|---|---|---|
| | Gains - Products | | | Losses - Products | | |
| | Product space | | | Product space | | |
| $E$ and $E_2$ | 0.162 | 0.364 | 0.503 | 0.094 | 0.258 | 0.381 |
| $D$ | 0.151 | 0.343 | 0.475 | 0.087 | 0.243 | 0.366 |
| $\widetilde{D}$ | **0.174** | **0.378** | **0.516** | **0.095** | **0.266** | **0.390** |
| | Country space | | | Country space | | |
| $E_2^*$ | 0.184 | 0.374 | 0.498 | 0.042 | 0.144 | 0.241 |
| $E^*$ | 0.126 | 0.299 | 0.432 | 0.079 | 0.231 | 0.346 |
| $D^*$ | 0.216 | 0.439 | 0.577 | 0.089 | 0.245 | 0.359 |
| $\widetilde{D}^*$ | 0.142 | 0.348 | 0.485 | **0.094** | **0.254** | **0.370** |
| | Combined space | | | Combined space | | |
| $E_2^{Tot}$ | 0.164 | 0.366 | 0.506 | 0.095 | 0.259 | 0.383 |
| $E^{Tot}$ | 0.165 | 0.368 | 0.506 | 0.094 | 0.259 | 0.382 |
| $D^{Tot}$ | 0.152 | 0.344 | 0.477 | 0.087 | 0.245 | 0.366 |
| $\widetilde{D}^{Tot}$ | **0.175** | **0.381** | **0.518** | 0.098 | 0.267 | 0.392 |
| | Gains - Countries | | | Losses - Countries | | |
| | Product space | | | Product space | | |
| $E_2$ | 0.741 | 0.827 | 0.863 | 0.325 | 0.443 | 0.510 |
| $E$ | 0.783 | 0.856 | 0.878 | 0.499 | 0.630 | 0.713 |
| $D$ | **0.970** | **0.986** | **0.992** | **0.746** | **0.848** | **0.887** |
| $\widetilde{D}$ | 0.758 | 0.829 | 0.867 | 0.616 | 0.771 | 0.829 |
| | Country space | | | Country space | | |
| $E_2^*$ and $E^*$ | 0.972 | 0.991 | 0.995 | 0.801 | 0.888 | 0.931 |
| $D^*$ | 0.873 | 0.926 | 0.942 | 0.658 | 0.767 | 0.822 |
| $\widetilde{D}^*$ | 0.974 | 0.992 | 0.994 | 0.814 | 0.898 | 0.931 |
| | Combined space | | | Combined space | | |
| $E_2^{Tot}$ | 0.773 | 0.869 | 0.906 | 0.381 | 0.523 | 0.589 |
| $E^{Tot}$ | 0.773 | 0.867 | 0.904 | 0.380 | 0.524 | 0.590 |
| $D^{Tot}$ | **0.967** | **0.988** | **0.992** | **0.767** | **0.860** | **0.897** |
| $\widetilde{D}^{Tot}$ | 0.765 | 0.831 | 0.869 | 0.620 | 0.779 | 0.834 |

**Table A2. Summary of test results for 2-period changes**

| | p ≤ 0.01 | p ≤ 0.05 | p ≤ 0.10 | p ≤ 0.01 | p ≤ 0.05 | p ≤ 0.10 |
|---|---|---|---|---|---|---|
| | Gains - Products | | | Losses - Products | | |
| | Product space | | | Product space | | |
| $E$ and $E_2$ | 0.277 | 0.467 | 0.591 | 0.126 | 0.289 | 0.411 |
| $D$ | 0.268 | 0.449 | 0.569 | 0.115 | 0.274 | 0.393 |
| $\widetilde{D}$ | **0.290** | **0.481** | **0.604** | **0.127** | 0.301 | **0.422** |
| | Country space | | | Country space | | |
| $E_2^*$ | 0.293 | 0.475 | 0.580 | 0.078 | 0.189 | 0.289 |
| $E^*$ | 0.238 | 0.408 | 0.527 | 0.106 | 0.257 | 0.374 |
| $D^*$ | 0.336 | 0.548 | 0.666 | **0.125** | 0.288 | 0.405 |
| $\widetilde{D}^*$ | 0.264 | 0.460 | 0.590 | 0.124 | **0.288** | **0.407** |
| | Combined space | | | Combined space | | |
| $E_2^{Tot}$ | 0.279 | 0.470 | 0.594 | 0.127 | 0.291 | 0.412 |
| $E^{Tot}$ | 0.280 | 0.470 | 0.594 | 0.126 | 0.292 | 0.412 |
| $D^{Tot}$ | 0.268 | 0.450 | 0.572 | 0.116 | 0.277 | 0.396 |
| $\widetilde{D}^{Tot}$ | **0.291** | **0.485** | **0.606** | 0.129 | **0.301** | 0.423 |
| | Gains - Countries | | | Losses - Countries | | |
| | Product space | | | Product space | | |
| $E_2$ | 0.800 | 0.862 | 0.895 | 0.358 | 0.464 | 0.545 |
| $E$ | 0.835 | 0.881 | 0.905 | 0.519 | 0.638 | 0.714 |
| $D$ | **0.985** | 0.997 | 0.997 | **0.783** | **0.868** | **0.910** |
| $\widetilde{D}$ | 0.785 | 0.844 | 0.877 | 0.653 | 0.788 | 0.853 |
| | Country space | | | Country space | | |
| $E_2^*$ and $E^*$ | 0.988 | **0.995** | 0.996 | 0.817 | 0.913 | 0.946 |
| $D^*$ | 0.906 | 0.941 | 0.956 | 0.665 | 0.778 | 0.827 |
| $\widetilde{D}^*$ | 0.990 | 0.994 | 0.997 | 0.840 | 0.926 | 0.950 |
| | Combined space | | | Combined space | | |
| $E_2^{Tot}$ | 0.846 | 0.914 | 0.932 | 0.431 | 0.542 | 0.603 |
| $E^{Tot}$ | 0.848 | 0.910 | 0.932 | 0.429 | 0.544 | 0.603 |
| $D^{Tot}$ | **0.987** | 0.997 | 0.997 | **0.794** | **0.888** | **0.919** |
| $\widetilde{D}^{Tot}$ | 0.785 | 0.844 | 0.880 | 0.665 | 0.799 | 0.866 |

**Table A3. Summary of test results for 3-period changes**

| | $p \leq 0.01$ | $p \leq 0.05$ | $p \leq 0.10$ | $p \leq 0.01$ | $p \leq 0.05$ | $p \leq 0.10$ |
|---|---|---|---|---|---|---|
| | Gains - Products | | | Losses - Products | | |
| | Product space | | | Product space | | |
| $E$ and $E_2$ | 0.214 | 0.426 | 0.569 | 0.110 | 0.273 | 0.393 |
| $D$ | 0.203 | 0.409 | 0.544 | 0.099 | 0.260 | 0.380 |
| $\widetilde{D}$ | **0.230** | **0.441** | **0.581** | **0.113** | **0.286** | **0.410** |
| | Country space | | | Country space | | |
| $E_2^*$ | 0.235 | 0.438 | 0.556 | 0.065 | 0.186 | 0.294 |
| $E^*$ | 0.170 | 0.364 | 0.495 | 0.087 | 0.239 | 0.352 |
| $D^*$ | 0.292 | 0.525 | 0.651 | 0.106 | **0.280** | **0.399** |
| $\widetilde{D}^*$ | 0.208 | 0.431 | 0.569 | **0.108** | 0.270 | 0.390 |
| | Combined space | | | Combined space | | |
| $E_2^{Tot}$ | 0.217 | 0.430 | 0.570 | 0.111 | 0.274 | 0.397 |
| $E^{Tot}$ | 0.217 | 0.430 | 0.569 | 0.110 | 0.277 | 0.397 |
| $D^{Tot}$ | 0.205 | 0.411 | 0.545 | 0.099 | 0.262 | 0.381 |
| $\widetilde{D}^{Tot}$ | **0.232** | **0.445** | **0.583** | 0.115 | 0.287 | 0.412 |
| | Gains - Countries | | | Losses - Countries | | |
| | Product space | | | Product space | | |
| $E_2$ | 0.819 | 0.882 | 0.898 | 0.371 | 0.493 | 0.557 |
| $E$ | 0.866 | 0.903 | 0.915 | 0.536 | 0.661 | 0.714 |
| $D$ | **0.994** | **0.997** | 0.998 | **0.807** | **0.892** | **0.927** |
| $\widetilde{D}$ | 0.816 | 0.879 | 0.897 | 0.710 | 0.820 | 0.860 |
| | Country space | | | Country space | | |
| $E_2^*$ and $E^*$ | 0.995 | **0.997** | 0.998 | 0.860 | 0.925 | 0.945 |
| $D^*$ | 0.918 | 0.944 | 0.965 | 0.687 | 0.792 | 0.820 |
| $\widetilde{D}^*$ | 0.995 | **0.997** | 0.997 | 0.862 | 0.930 | 0.952 |
| | Combined space | | | Combined space | | |
| $E_2^{Tot}$ | 0.874 | 0.911 | 0.942 | 0.456 | 0.569 | 0.616 |
| $E^{Tot}$ | 0.874 | 0.911 | 0.940 | 0.459 | 0.567 | 0.617 |
| $D^{Tot}$ | **0.994** | 0.998 | 0.998 | **0.819** | **0.895** | **0.932** |
| $\widetilde{D}^{Tot}$ | 0.821 | 0.881 | 0.902 | 0.715 | 0.819 | 0.864 |

**Table A4. Summary of test results for 4-period changes**

| | $p \leq 0.01$ | $p \leq 0.05$ | $p \leq 0.10$ | $p \leq 0.01$ | $p \leq 0.05$ | $p \leq 0.10$ |
|---|---|---|---|---|---|---|
| | Gains - Products | | | Losses - Products | | |
| | Product space | | | Product space | | |
| $E$ and $E_2$ | 0.230 | 0.446 | 0.577 | 0.107 | 0.270 | 0.394 |
| $D$ | 0.219 | 0.433 | 0.559 | 0.097 | 0.256 | 0.378 |
| $\widetilde{D}$ | **0.246** | **0.463** | **0.594** | **0.112** | **0.288** | **0.405** |
| | Country space | | | Country space | | |
| $E_2^*$ | 0.256 | 0.453 | 0.570 | 0.071 | 0.192 | 0.300 |
| $E^*$ | 0.185 | 0.384 | 0.514 | 0.087 | 0.232 | 0.354 |
| $D^*$ | 0.313 | 0.549 | 0.673 | **0.107** | **0.285** | **0.410** |
| $\widetilde{D}^*$ | 0.228 | 0.461 | 0.596 | 0.101 | 0.267 | 0.387 |
| | Combined space | | | Combined space | | |
| $E_2^{Tot}$ | 0.232 | 0.450 | 0.580 | 0.107 | 0.272 | 0.397 |
| $E^{Tot}$ | 0.233 | 0.450 | 0.582 | 0.107 | 0.273 | 0.397 |
| $D^{Tot}$ | 0.222 | 0.436 | 0.561 | 0.099 | 0.259 | 0.380 |
| $\widetilde{D}^{Tot}$ | **0.250** | **0.467** | **0.596** | 0.112 | 0.288 | 0.409 |
| | Gains - Countries | | | Losses - Countries | | |
| | Product space | | | Product space | | |
| $E_2$ | 0.830 | 0.888 | 0.914 | 0.384 | 0.476 | 0.538 |
| $E$ | 0.884 | 0.899 | 0.916 | 0.536 | 0.680 | 0.716 |
| $D$ | **0.994** | 0.998 | 0.998 | **0.809** | **0.889** | **0.924** |
| $\widetilde{D}$ | 0.843 | 0.880 | 0.899 | 0.687 | 0.813 | 0.862 |
| | Country space | | | Country space | | |
| $E_2^*$ and $E^*$ | 0.998 | 0.998 | 0.998 | 0.853 | 0.920 | 0.944 |
| $D^*$ | 0.920 | 0.948 | 0.955 | 0.673 | 0.791 | 0.824 |
| $\widetilde{D}^*$ | 0.998 | 0.998 | 0.998 | 0.860 | 0.924 | 0.942 |
| | Combined space | | | Combined space | | |
| $E_2^{Tot}$ | 0.873 | 0.925 | 0.953 | 0.460 | 0.562 | 0.618 |
| $E^{Tot}$ | 0.875 | 0.925 | 0.953 | 0.453 | 0.562 | 0.618 |
| $D^{Tot}$ | **0.994** | 0.998 | 0.998 | **0.811** | **0.900** | **0.936** |
| $\widetilde{D}^{Tot}$ | 0.843 | 0.877 | 0.901 | 0.693 | 0.818 | 0.864 |

**Table A5. Summary of test results for 5-period changes**

| | p ≤ 0.01 | p ≤ 0.05 | p ≤ 0.10 | p ≤ 0.01 | p ≤ 0.05 | p ≤ 0.10 |
|---|---|---|---|---|---|---|
| | Gains - Products | | | Losses - Products | | |
| | Product space | | | Product space | | |
| $E$ and $E_2$ | 0.253 | 0.478 | 0.608 | 0.111 | 0.275 | 0.400 |
| $D$ | 0.245 | 0.466 | 0.591 | 0.104 | 0.261 | 0.383 |
| $\widetilde{D}$ | **0.271** | **0.494** | **0.623** | 0.119 | **0.289** | **0.414** |
| | Country space | | | Country space | | |
| $E_2^*$ | 0.275 | 0.473 | 0.587 | 0.078 | 0.204 | 0.308 |
| $E^*$ | 0.209 | 0.415 | 0.547 | 0.093 | 0.243 | 0.358 |
| $D^*$ | 0.346 | 0.584 | 0.704 | **0.112** | **0.289** | **0.407** |
| $\widetilde{D}^*$ | 0.264 | 0.498 | 0.633 | 0.107 | 0.263 | 0.383 |
| | Combined space | | | Combined space | | |
| $E_2^{Tot}$ | 0.255 | 0.479 | 0.611 | 0.110 | 0.277 | 0.402 |
| $E^{Tot}$ | 0.256 | 0.480 | 0.611 | 0.112 | 0.276 | 0.402 |
| $D^{Tot}$ | 0.248 | 0.467 | 0.592 | 0.105 | 0.262 | 0.384 |
| $\widetilde{D}^{Tot}$ | **0.275** | **0.498** | **0.626** | **0.119** | 0.293 | 0.416 |
| | Gains - Countries | | | Losses - Countries | | |
| | Product space | | | Product space | | |
| $E_2$ | 0.845 | 0.887 | 0.919 | 0.393 | 0.503 | 0.567 |
| $E$ | 0.871 | 0.906 | 0.916 | 0.543 | 0.640 | 0.700 |
| $D$ | 0.994 | 0.997 | 0.997 | **0.793** | **0.877** | **0.900** |
| $\widetilde{D}$ | 0.832 | 0.877 | 0.897 | 0.683 | 0.820 | 0.850 |
| | Country space | | | Country space | | |
| $E_2^*$ and $E^*$ | 0.994 | **0.994** | 0.997 | 0.860 | 0.923 | 0.947 |
| $D^*$ | 0.913 | 0.932 | 0.952 | 0.683 | 0.773 | 0.833 |
| $\widetilde{D}^*$ | 0.994 | **0.994** | 0.997 | 0.860 | 0.923 | 0.950 |
| | Combined space | | | Combined space | | |
| $E_2^{Tot}$ | 0.894 | 0.929 | 0.945 | 0.477 | 0.557 | 0.623 |
| $E^{Tot}$ | 0.894 | 0.929 | 0.948 | 0.470 | 0.563 | 0.623 |
| $D^{Tot}$ | 0.994 | 0.997 | 0.997 | **0.817** | **0.883** | **0.903** |
| $\widetilde{D}^{Tot}$ | 0.835 | 0.877 | 0.897 | 0.687 | 0.827 | 0.850 |

**Table A6. Summary of test results for 6-period changes**

| | p ≤ 0.01 | p ≤ 0.05 | p ≤ 0.10 | p ≤ 0.01 | p ≤ 0.05 | p ≤ 0.10 |
|---|---|---|---|---|---|---|
| | Gains - Products | | | Losses - Products | | |
| | Product space | | | Product space | | |
| $E$ and $E_2$ | 0.292 | 0.516 | 0.635 | 0.114 | 0.295 | 0.418 |
| $D$ | 0.288 | 0.500 | 0.620 | 0.112 | 0.286 | 0.417 |
| $\widetilde{D}$ | **0.308** | **0.533** | **0.650** | **0.124** | **0.313** | **0.447** |
| | Country space | | | Country space | | |
| $E_2^*$ | 0.296 | 0.489 | 0.597 | 0.090 | 0.222 | 0.330 |
| $E^*$ | 0.246 | 0.454 | 0.580 | 0.097 | 0.257 | 0.385 |
| $D^*$ | 0.382 | 0.623 | 0.735 | **0.114** | **0.288** | **0.416** |
| $\widetilde{D}^*$ | 0.297 | 0.539 | 0.665 | 0.098 | 0.262 | 0.381 |
| | Combined space | | | Combined space | | |
| $E_2^{Tot}$ | 0.295 | 0.520 | 0.638 | 0.118 | 0.299 | 0.422 |
| $E^{Tot}$ | 0.292 | 0.518 | 0.638 | 0.117 | 0.298 | 0.424 |
| $D^{Tot}$ | 0.290 | 0.504 | 0.622 | 0.112 | 0.284 | 0.418 |
| $\widetilde{D}^{Tot}$ | **0.314** | **0.534** | **0.653** | 0.125 | 0.309 | 0.449 |
| | Gains - Countries | | | Losses - Countries | | |
| | Product space | | | Product space | | |
| $E_2$ | 0.865 | 0.923 | 0.935 | 0.409 | 0.523 | 0.570 |
| $E$ | 0.884 | 0.916 | 0.916 | 0.517 | 0.671 | 0.718 |
| $D$ | 0.994 | 0.994 | 0.994 | **0.792** | **0.879** | **0.919** |
| $\widetilde{D}$ | 0.852 | 0.871 | 0.884 | 0.705 | 0.805 | 0.832 |
| | Country space | | | Country space | | |
| $E_2^*$ and $E^*$ | 0.994 | 0.994 | 0.994 | 0.846 | 0.906 | 0.940 |
| $D^*$ | 0.923 | 0.935 | 0.942 | 0.685 | 0.765 | 0.812 |
| $\widetilde{D}^*$ | 0.994 | 0.994 | 0.994 | 0.866 | 0.906 | 0.946 |
| | Combined space | | | Combined space | | |
| $E_2^{Tot}$ | 0.916 | 0.942 | 0.961 | 0.483 | 0.597 | 0.644 |
| $E^{Tot}$ | 0.923 | 0.942 | 0.961 | 0.490 | 0.591 | 0.638 |
| $D^{Tot}$ | **0.994** | 0.994 | 0.994 | **0.805** | **0.893** | **0.919** |
| $\widetilde{D}^{Tot}$ | 0.852 | 0.871 | 0.884 | 0.691 | 0.805 | 0.832 |